\documentclass[10pt]{article}
\usepackage{amsmath,amsfonts,amssymb,amscd,amsxtra,amsthm}

\usepackage{graphicx}
\usepackage{epsfig}
\usepackage{bm}
\begin{document}
\title{Constrains on treating linear part as perturbation in a QCD Potential Model}
\author{ $^{1}$ D K Choudhury and  $^{2}$Krishna Kingkar Pathak \\
$^{1}$ Deptt. of Physics, Gauhati University, Guwahati-781014,India\\
$^{2}$ Deptt. of Physics,Arya Vidyapeeth College,Guwahati-781016,India
e-mail:kkingkar@gmail.com}
\date{}
\maketitle

\begin{abstract}
We make a review of the different works of Quark Model which are based on Perturbation theory. We make a critical analysis of taking linear part of the Potential as perturbation with Coulombic part as parent and its reverse case. We find in the analysis that the linear part can be treated as perturbation for a set of larger values of $\alpha_{s}$ in the range $0.37 \leq \alpha_{s}\leq 0.75$ with a constant shift in the Cornell potential within range of $-0.4 GeV \leq c \leq -1 GeV$. Moreover with the same range of constant shift in the Potential, we expect better results with coulombic part as perturbation for $\alpha_{s}\leq 0.37$.
 
\end{abstract} 
\section{ Introduction}

 The theory of QCD with just one heavy quark is found to be quite useful in phenomenological Models for investigation of hadrons at the hadronic scale. In the Potential Models, the effective Potential between a quark and antiquark can be taken as the Coulomb-plus-linear potential. The form of the Potential
 \begin{equation} 
  V (r) = -a/r + br+c,
  \end{equation}
   which is also known as the Cornell potential, has received a great deal of attention both in particle physics, more precisely in the context of meson spectroscopy where it is used to describe systems of quark and antiquark bound states, and in atomic and molecular physics where it represents a radial Stark effect in hydrogen. This potential was used with considerable success in models describing systems of bound heavy quarks. The potential includes the short distance Coulombic interaction of quarks mediated by a single gluon exchange, known from perturbative quantum chromodynamics (QCD), and the large distance confinement that arises from the color field flux tube between the quarks, known from lattice QCD, via the linear term in a simple form. Coulombic term alone is not sufficient because it would allow free quarks to ionize from the system. \\
The coefficients $a$ and $b$ are adjusted to fit the charmonium spectrum, but with the assumption that, it should be valid for all other heavy quarkonia. As such, the flavour dependence should arise solely from the mass of the bound quarks. However it has been found to be questionable about the numbers of free parameters and numbers of findings in any Potential Model. The success of a Phenomenological Model depends on reducing the free model parameters to obtain more precise values with proper arguments and analysis. 
  we put forward the comments on linear part of the Potential as perturbation with coulombic part as Parent\cite{7,9} as well as coulombic part as perturbation with linear as parent \cite{10} in a specific Potential Model and attempt to put some cnstraints on the model parameters with some analysis.

 \section{The method of Perturbation}
 It is well known that one cannot solve the Schrödinger equation in quantum mechanics with the QCD potential(eq.1) except for some simple models. For that reason many authors have devoted considerable time and effort
to develop efficient approximate methods. Among them, perturbation theory has been helpful since
the earliest applications of quantum mechanics. One of the main advantages of this approach is that
it provides analytical approximate solutions for many nontrivial simple problems which are suitable
for subsequent discussion and interpretation of the physical phenomena. In fact, perturbation theory
is probably one of the approximate methods that most appeals to intuition \cite{3}.\\ 
While dealing with perturbation theory, one has to check convergence of the series appearing in the procedure. If the rate of convergence of the perturbation series is sufficiently great, we may obtain accurate
results without difficulty by summing all available terms. If, on the other hand, the series is divergent
or slowly convergent we may need an appropriate summation algorithm to obtain acceptable results.
For concreteness and simplicity we focus on the perturbation series for the energy. The test of convergence of power series is being discussed in chapter.6 of ref.\cite{3}.\\

In this report, we briefly comment on some of the results and conclusions
derived by us and other authors keeping present contribution as simple as possible. Moreover, we mainly
concentrate on the practical aspect of obtaining accurate results from the perturbation series. In conventional perturbation theory, the Hamiltonian H is written as $H_{0}+H^{\prime}$ , where the
eigenvalues and eigenstates of $H_{0}$ are known.\\
The advantage of taking Cornell Potential for study is that it leads naturally to two choices of "parent" Hamiltonian, one based on the Coulomb part and the other on the linear term, which can be usefully compared. It is expected that a critical role is played by $ r_{0} $ where the Potential $ V(r)=0 $. Aitchison and Dudek 
in ref.[4] put an argument that if the size of a state measured by $\langle r\rangle < r_{0}$, then the Coulomb part as the "Parent" will perform better and if not so the linear part as "parent" will perform better. The Aitchison's work also showed the results that with Coumbic part as perturbation(VIPT), bottomonium spectra are well explained than Charmonium where as Charmonium states are well explained with linear part as parent. It becomes noteworthy in this context that the critical distance $ r_{0} $ is not a constant and can be enhanced by reducing $b$ and $c$ or by increasing $\alpha_{s}$. In fig.1 we show the variation of $V(r)$ with the variation of $b$ and $c$.\\
\begin{figure}
\begin{center}
\includegraphics[scale=0.8]{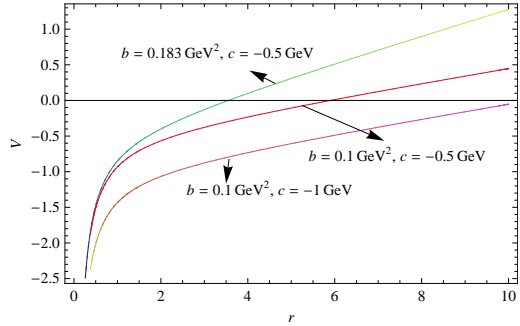}  
\caption {Variation of $V(r)$ with $b$ and $c$.}
\end{center}
\end{figure}
\section{The QCD Potential Model}
The QCD Potential Model to be discussed here has its origin of the work of De Rujula etal.\cite{5}. Since then it has been modified from time to time. For completeness and proper reference we put the last modified version of the wavefunction with coulombic part as parent as\cite{9}
\begin{equation}
\psi_{rel+conf}\left(r\right)=\frac{N^{\prime}}{\sqrt{\pi a_{0}^{3}}} e^{\frac{-r}{a_{0}}}\left( C^{\prime}-\frac{\mu b a_{0} r^{2}}{2}\right)\left(\frac{r}{a_{0}}\right)^{-\epsilon}.
\end{equation}
where the terms are explained in the ref.\cite{9}
\begin{equation}
\epsilon=1-\sqrt{1-\left(\frac{4}{3}\alpha_{s}\right)^{2}}.
\end{equation}
 Question arises about the validity and consequences of treating linear confinement potential as perturbation as the confining potential is expected to be more dominant then coulombic piece. However the confinement potential is operative within the charateristic size of a hadron. The characteristic distance of heavy flavour mesons is taken to be of the order of $ \frac{m_{Q}}{2} $,$m_{Q}$ being the mass of the heavy quark\cite{11}. For heavy-light meson even the corresponding distance is smaller than the hadronic scale $\Lambda^{-1}_{QCD}$. A similar observation has been done in QCD Sum rule approach\cite{12}. Considering such observation seriously, one can treat linear confining pat as perturbation for mesons containing at least one heavy flavour.\\
 \section{Constrains from two points of view}
 The values of $ \alpha_{s}$ and the constant shift of the potential $V(r)$ are found to vary from model to model. It is expected to fit from mass spectroscopy of hadrons to study its other  properties. A narrow range of the free parameters in a Potential Model measures its success and applicability as well. Here we have tried to show some constrains on the free parameters $\alpha_{s}$ and $c$ from two points of view.
\subsection{From the convergence Point of view} 
From the momentum transform of eq.2 \cite{13}, we see that for a lower cut-off value of $Q^{2}_{0}$, either one has to consider a very small value of $b$ or to increase the value of $\alpha_{s}$, which is obvious, since in both the cases coulombic part will be more dominant(eq.1). However reality condition of $\epsilon$ demands that $\alpha_{s} \leq \frac{3}{4}$ and hence one cant go beyond $\alpha_{s}=0.75$ in this approach. The standard spectroscopic result $b=0.183 GeV^{2}$\cite{14} can be accomodated by a proper choice of $c$ such that the perturbative condition of eq.5 becomes
 \begin{equation}
 \frac{(4-\epsilon)(3-\epsilon)\mu b a^{3}_{0}}{2(1+a^{2}_{0}Q^{2})}<<C^{\prime}.
\end{equation}
This possibility arises due to the arbitrariness of $A_{0}$, which appears in the series solution. With this one can impose $b=0.183 GeV^{2}$ with low $Q^{2}$ value. 
The  values of $Q^{2}_{0}$ with $ b=0.183 GeV^{2} $for $ B $ and$ D $ mesons are shown in table.1.
\begin{table}
\begin{center}
\caption{ Values of $Q^{2}_{0}$ (in $GeV^{2}$) for heavy-light mesons with linear part as perturbation and $ m_{u/d}=0.33 $GeV,$ m_{s}=0.483 $GeV,$ m_{c}=1.55 $GeV,$ m_{b}=4.93 $GeV }
\begin{tabular}{|c|c|c|c|c|}
\hline Mesons & $\alpha_{s}=0.65$  & $\alpha_{s}=0.5$  & $\alpha_{s}=0.4$& $\alpha_{s}=0.37$ \\ 
\hline $D$ & -0.0001 & 0.0264 & 0.0362&0.037 \\ 
\hline $D_{s}$ &-0.0166  &0.0304  & 0.0490& 0.052\\ 
\hline  $B$&-0.0057  & 0.0286 & 0.0412 &0.043\\ 
\hline $B_{s}$ &-0.0371  & 0.0300 & 0.0530 &0.056\\ 
\hline $B_{c}$ &-0.7045  &-0.0219  &0.0162 & 0.054\\ 
\hline 
\end{tabular} 
\end{center}
\end{table}
Thus to incorporate lower value of $Q^{2}$ ($Q^{2}\leq\Lambda_{QCD}^{2}$),with linear part as perturbation in the improved version one expects a bound of $\alpha_{s}$ as $0.37 \leq \alpha_{s}\leq 0.75$.\\

\subsection{From the condition of Aitchison and Dudek}

Considering the argument of Aitchison and Dudek \cite{4} $\langle r\rangle < r_{0}$ to treat the linear part as perturbation, 
we get 
\begin{equation}
\langle r\rangle_{coul} = \int  \psi^{*} r \psi dr =\frac {3a_{0}} {2}=r_{1}(say)
\end{equation}
and the critical distance $r_{0}$ at which $V(r_{0})=0$ can be obtained by the relation \\
\begin{equation}
br_{0}^{2} +cr_{0}-\frac{4\alpha_{s}}{3}=0.
\end{equation}
The variation of $r_{1}$ and $r_{0}$ with the model parameters can be easily studied from the above relations and the results are tabulated in table.2. From the results it seems to be clear that to treat linear part as perturbation with the valid condition $\langle r\rangle < r_{0}$ one has to choose the value of $c\ll-0.5$. With $c=-1 GeV$ the condition is found to be valid for certain value of $\alpha_{s}$.

\begin{table}
\begin{center}
\caption{ Values of $r_{1}$ and $ r_{0}$ for different mesons with $m_{u/d}=0.33 GeV$,$ m_{s}=0.483 GeV,$ $m_{c}=1.55 GeV,$ $m_{b}=4.97 GeV,$ $b=0.183 GeV^{2}$ and $ c=-1GeV$}  
\begin{tabular}{|c|c|c|c|c|c|c|}\hline
mesons & \multicolumn{2}{c|}{ $\alpha_{s}=0.65$} & \multicolumn{2}{c|}{$\alpha_{s}=0.5$} & \multicolumn{2}{c|}{$\alpha_{s}=0.4$}   \\\hline 
 &\multicolumn{1}{l|}{$r_{1}$}& $r_{0}$ &\multicolumn{1}{l|}{$r_{1}$}&$r_{0}$&\multicolumn{1}{l|}{$r_{1}$}&$r_{0}$\\\hline
$D(c\bar{u}/c\bar{d})$&\multicolumn{1}{l|}{6.26}&6.22&\multicolumn{1}{l|}{8.14}&6.06&\multicolumn{1}{l|} {10.18}&5.95\\\hline
$D(c\bar{s})$&\multicolumn{1}{l|}{4.70}&6.22&\multicolumn{1}{l|}{6.11}&6.06&\multicolumn{1}{l|}{7.63}&5.95\\\hline
$B(\bar{b}u/\bar{b}d)$&\multicolumn{1}{l|}{5.50}&6.22&\multicolumn{1}{l|}{7.15}&{6.06}&\multicolumn{1}{l|}{8.93}&5.95\\\hline
$B_{s}(\bar{b}s)$&\multicolumn{1}{l|}{3.93}&6.22&\multicolumn{1}{l|}{5.1}&{6.06}&\multicolumn{1}{l|}{6.39
}&5.95\\\hline
$B_{c}(\bar{b}c)$&\multicolumn{1}{l|}{1.46}&6.22&\multicolumn{1}{l|}{1.90}&{6.06}&\multicolumn{1}{l|}{2.38}&5.95\\\hline

\end{tabular}
\end{center}
\end{table}

\subsection{Constrains on $\alpha_{s}$}
From the above analysis we see that in the perturbation procedure the value of $\alpha_{s}$ and the model parameter $c$ 
 plays a crucial role in choosing the parent and perturbative terms. From the reality condition and Convergency of series  demands the value of $\alpha_{s}$ within the range of $0.37 \leq \alpha_{s}\leq 0.75$ in the model without putting any further restriction or constraints.  However the logarithmic decrease of $\alpha_{s}$ depends on the QCD energy scale parameter $\Lambda_{QCD}$ which is a free parameter and has to be measured  in the experiments. One well known formula to fix the value of $\alpha_{s}$ in Quark models is taken as \cite{9}
  \begin{equation}
\alpha_{s}\left(\mu^{2}\right)=\frac{4\pi}{\left(11-\frac{2n_{f}}{3}\right)ln\left(\frac{\mu^{2}+M^{2}_{B}}{\Lambda_{QCD}^{2}}\right)} 
  \end{equation}
where, $n_{f}$ is the number of light flavours, $\mu$ is renormalistion scale related to the constituent quark masses as $\mu=2\frac{m_{i}m_{j}}{m_{i}+m_{j}}$ . $M_{B}$ is the background mass related to the confinement term of the potential as $M_{B}=2.24 \times b^{1/2}=0.95 GeV$.\\
The reality condition of $\alpha_{s}$ in eq.8 requires that $\Lambda_{QCD} \leq 460 MeV$.  By fitting the $\rho$ meson mass in eq.8 one easily obtains QCD scale parameter $\Lambda_{QCD}=413 MeV$. 
  With this energy scale parameter we can obtain the values of $\alpha_{s}$ for the heavy-light mesons  such as $\alpha_{s}(D)=0.71$,$\alpha_{s}(D_{s})=0.65$,$\alpha_{s}(B)=0.69$,$\alpha_{s}(B_{s})=0.61$ and $\alpha_{s}(B_{c})=0.38$.
  
 \section{Conclusion and Comments}  
  In this letter we mainly devote in finding the analytical conditions to treat the linear part of the Cornell potential as perturbation.  Morever the findings in the analysis can be conluded as:
  \begin{itemize}
\item From the convergence point of view one can consider the confinig part of the potential as perturbation with $0.4 \leq \alpha_{s}\leq 0.75$ and $c=-0.5 GeV$.

\item The validity of the condition $\langle r\rangle < r_{0}$ demands parametrisation of $\langle c\rangle < -0.5 GeV$ and $\alpha_{s}>0.6$.
\item If the above two points are given the equal footings then the renormalization scale of the model should be  set to be $\Lambda_{QCD}=413$ MeV.

\end{itemize}

 However  with linear part as perturbation, if the value of $\alpha_{s}$ in the above range is taken to be granted, then with the same potential another possibility of considering the coulombic part as perturbation also arises for a value of $\alpha_{s}\leq 0.37$. Interestingly, in ref.\cite{10}, it is shown that with $\alpha_{s}=0.39$ and $\alpha_{s}=0.22$ one can obtain the required values of slope and curvature in the model with coulombic part as perturbation. The results in ref.\cite{10} clearly indicates that with coulombic part as perturbation, one can get improved results with $\alpha_{s} \leq 0.37$ than  $\alpha_{s} \geq 0.37$. \\

\end{document}